\documentclass [pra,twocolumn,preprintnumbers,amsmath,amssymb]{revtex4}

\usepackage{graphicx}
\usepackage{dcolumn}
\usepackage{bm}

\DeclareGraphicsExtensions{.eps,.gif,.ps}

\begin{document}

\title{Mutually Unbiased Bases and Complementary Spin 1 Observables}

\author{Pawe{\l} Kurzy\'nski}
\email{kurzpaw@hoth.amu.edu.pl}

\author{Wawrzyniec Kaszub}

\author{Miko{\l}aj Czechlewski}

\affiliation{Faculty of Physics, Adam Mickiewicz University,
Umultowska 85, 61-614 Pozna\'{n}, Poland.}


\begin{abstract}
The two observables are complementary if they cannot be measured simultaneously, however they become {\it maximally} complementary if their eigenstates are mutually unbiased. Only then the measurement of one observable gives no information about the other observable. The spin projection operators onto three mutually orthogonal directions are maximally complementary only for the spin 1/2. For the higher spin numbers they are no longer unbiased. In this work we examine the properties of spin 1 Mutually Unbiased Bases (MUBs) and look for the physical meaning of the corresponding operators. We show that if the computational basis is chosen to be the eigenbasis of the spin projection operator onto some direction $z$, the states of the other MUBs have to be squeezed. Then, we introduce the analogs of momentum and position operators and interpret what information about the spin vector the observer gains while measuring them. Finally, we study the generation and the measurement of MUBs states by introducing the Fourier like transform through spin squeezing. The higher spin numbers are also considered.
\end{abstract}

\maketitle

\section{Introduction}

Two $d$ dimensional orthonormal bases $\{|a_j\rangle\}$ and $\{|b_k\rangle\}$ are called unbiased if 
\begin{equation}\label{e1}
\forall_{j,k}~~|\langle a_j|b_k\rangle|^2=\frac{1}{d},
\end{equation}
and the set of more than two bases of this kind is called mutually unbiased if the above holds for every pair of bases from this set. In quantum mechanics it means that if the two observables $A$ and $B$ have unbiased eigenbases the measurement of the observable $A$ reveals no information about possible outcomes of the measurement of the observable $B$ and vice versa.  

The study on Mutually Unbiased Bases (MUBs) has been started by Schwinger \cite{Schw} almost fifty years ago. The properties of MUBs have been successfully applied in many areas of quantum physics: they provide the security of quantum key distribution protocols \cite{BB84,PP} and the solution to the Mean King problem \cite{VAA,AE1,AE2}, minimize the number of measurements needed to determine the quantum state \cite{Iv,WF} and are related to the discrete Wigner Functions \cite{GHW}. Despite their wide use, there is still an intriguing open question about the maximal number of such bases in the non-prime power dimensional complex spaces \cite{Prob}. It is well known that the MUBs have deep physical meaning for the quantum systems described by the continuous Hilbert spaces. The eigenstates of the position and the momentum operators are mutually unbiased. The same stands for the electric and the magnetic field operators. In the case of spin, the Hilbert space is finite dimensional. The three MUBs of spin $1/2$ have nice geometrical representation, because they are simply the eigenbases of the spin projection operators onto any three mutually orthogonal directions. This is not the case for the higher spin numbers $S>1/2$. The Hilbert space of spin $1/2$ is isomorphic with the three dimensional real space what is the essence of the Bloch sphere picture. On the other hand, the Hilbert space of spin $S>1/2$ is much richer than the three dimensional real space, that is why it would be naive to expect that the MUBs of higher spins have simple graphical interpretation. 

Most of the research done on MUBs are mainly based on the mathematical properties of the underlying Hilbert space and the intuitive physical picture behind the complex state vectors is somehow lost. We are trying to bring back this picture by studying the properties of MUBs for spin 1 states. What motivates our work, is the fact that the knowledge of the observables corresponding to the MUBs and the ability to generate and measure unbiased states for a certain system is necessary to fully exploit it in quantum information processing tasks. In the following paper we find that for the computational basis being the basis of projection operator onto any direction, the remaining MUBs have to be squeezed. We give the physical interpretation of the two operators which are maximally complementary, one of them being the projection operator. They are related by the Fourier transform and might be considered as an analog of the position and the momentum. Then, we show how to transform between different MUBs by introducing the Fourier like transform through spin squeezing. Finally, the eigenstates of projection operator and their Fourier transforms are also briefly examined for the higher spin numbers. In the end, we discuss our results in the context of the recent studies on the biphoton implementation of a qutrit \cite{Bog1,Bog2,Lan,Bre}.

\section{Spin 1 states unbiased to the $\mathbf{S_z}$ basis}

The most common choice of the computational basis for spin 1 is the eigenbasis of the spin projection operator onto some direction $z$. Any state which is unbiased to all the states from the computational basis is of the form 
\begin{equation}\label{e2}
\frac{1}{\sqrt{3}}\begin{pmatrix}
             1 \\
             e^{i\alpha} \\
				 e^{i\beta} \\
				 \end{pmatrix}
\end{equation}
where $\alpha$ and $\beta$ are arbitrary phases. The first property of the above state is the zero mean value of the $z$ component of the mean spin vector $\langle \vec{S} \rangle=(\langle S_x \rangle, \langle S_y \rangle, \langle S_z \rangle )$, where $S_i$'s are the spin projection operators onto direction $i$ obeying the cyclic commutation relation $[S_i,S_j]=i\varepsilon_{ijk}S_k$. This is because the operator $S_z$ is diagonal, its eigenvalues are 1, 0 and -1 (here and throughout the work we take $\hbar=1$) and for the state (\ref{e2}) 
\begin{equation}\label{e3}
\langle S_z \rangle = \frac{1}{3}\text{Tr}\{S_z\}=0.
\end{equation}
This means that either the mean spin vector is $\vec{0}$ or it lies in the $xy$ plane. Indeed, the coordinates of the mean spin vector are 
\begin{eqnarray}\label{e4}
\langle S_x \rangle & = & \sqrt{2}/3~ \left(\cos\alpha+\cos(\beta-\alpha)\right); \nonumber \\ 
\langle S_y \rangle & = & \sqrt{2}/3~ 
\left(\sin\alpha+\sin(\beta-\alpha)\right); \\
\langle S_z \rangle & = & 0. \nonumber
\end{eqnarray}
Note, that due to the rotational symmetry for the study of the physical properties of states (\ref{e2}) we can only consider the subclass of states for which $\langle S_y \rangle = 0$. The rest of the states might be generated by the rotation in the $xy$ plane. It is easy to see that the subclass $\langle S_y \rangle = 0$ corresponds to $\beta=0$, so the Eq. (\ref{e2}) becomes
\begin{equation}\label{e5}
\frac{1}{\sqrt{3}}\begin{pmatrix}
             1 \\
             e^{i\alpha} \\
				 1 \\
				 \end{pmatrix},
\end{equation}
or to $\beta=2\alpha-\pi$, what gives
\begin{equation}\label{e6}
\frac{1}{\sqrt{3}}\begin{pmatrix}
             1 \\
             e^{i\alpha} \\
				 -e^{i2\alpha} \\
				 \end{pmatrix}.
\end{equation}

We obtained the class of states unbiased to the computational basis which are parameterized only by $\alpha$. They are the states with the mean spin vector of length $|(2\sqrt{2}/3)\cos\alpha|$ pointing in $x$ direction (\ref{e5}), or a completely unpolarized states with $\langle \vec{S} \rangle=\vec{0}$ (\ref{e6}). Since the maximum value of the mean spin vector length is $2\sqrt{2}/3 < 1$, any state of the form (\ref{e2}), including (\ref{e5}) and (\ref{e6}), cannot be a coherent spin state --- the eigenstate of the spin projection operator onto any direction with the eigenvalue $\pm 1$. The observables with all the eigenstates of the form (\ref{e2}) have to be much more sophisticated than simply the spin projection operators. On the other hand, since the states (\ref{e6}) are completely unpolarized, it may happen that at least some of them correspond to the eigenstates of the spin projection operators with the eigenvalue $0$ --- the null projection states.   

Now, let us consider the uncertainties $\Delta S_x^2$, $\Delta S_y^2$ and $\Delta S_z^2$. The uncertainty relation for the spin projection operators onto any three mutually orthogonal directions yields: 
\begin{equation}\label{e7}
\Delta S_i^2 \Delta S_j^2 \geq \frac{1}{4}|\langle S_k \rangle|^2.
\end{equation} 
The above inequality is very sensitive to the choice of the directions $i$, $j$ and $k$ and it was shown \cite{KU} that it should be applied for the $k$ lying along the mean spin vector. With properly defined uncertainty relation the squeezed spin state is defined as the one for which the uncertainty $\Delta S_i^2$ of the spin projection operator onto direction orthogonal to the mean spin vector is smaller than $S/2$, what was done in \cite{KU}. In our case the state is squeezed if there exist a direction orthogonal to the mean spin vector for which $\Delta S_i^2 < 1/2$. The mean spin vector of the states (\ref{e5}) lies along $x$ direction, therefore $\Delta S_y^2=\langle S_y^2 \rangle$ and $\Delta S_z^2=\langle S_z^2 \rangle$. What is interesting, both uncertainties do not depend on $\alpha$
\begin{eqnarray}\label{e8}
\Delta S_y^2 & = & 1/3 \nonumber; \\
\Delta S_z^2 & = & 2/3, 
\end{eqnarray}
thus according to the definition the states (\ref{e5}) are squeezed in $y$ direction. The states (\ref{e6}) have zero mean spin vector and it is hard to choose the proper $k$ direction for the relation (\ref{e7}), however in this case the variances are
\begin{eqnarray}\label{e9}
\Delta S_y^2 & = & 2/3 + 1/3~\cos 2\alpha \nonumber; \\
\Delta S_z^2 & = & 2/3, 
\end{eqnarray}
and together with $\Delta S_x^2  =  2/3 - 1/3~\cos 2\alpha$ one may find that these states are the null projection states onto direction 
\begin{equation}\label{e10}
\vec{n}=1/\sqrt{3} \left(\sqrt{2} \cos\alpha, \sqrt{2} \sin\alpha, 1\right).
\end{equation}
This direction makes the tetrahedral angle $\frac{1}{2}\arctan \sqrt{2}\approx 54.7^o$ with $z$ axis. In the light of the recent Bell-like inequality for spin 1 \cite{Kl} and due to the maximal entanglement of two spin 1/2 particles forming up a spin 1 particle, the null projection states are considered to be the most non-classical ones. Moreover, these states might be viewed as a maximally squeezed states in the direction for which $\Delta S_i^2=0$, because in the case of spin 1 they may emerge during the squeezing of a coherent state as an intermediate states between the two coherent ones. 

\section{Spin 1 MUBs}

The four MUBs in the three dimensional complex space are most commonly known to be the eigenvectors of the four unitary operators from the Weyl-Heisenberg group: $U$, $V$, $UV$ and $UV^2$. All four operators have eigenvalues $1$, $q$, and $q^2$ and obey the relation $AB=qBA$, where $q$ is the third root of unity. The bases, up to normalization, are given by 
\begin{eqnarray}
& \begin{pmatrix} 1 \\ 0 \\ 0 \\ \end{pmatrix}, & ~~\begin{pmatrix} 0 \\ 1 \\ 0 \\ \end{pmatrix},~~~~\begin{pmatrix} 0 \\ 0 \\ 1 \\ \end{pmatrix} \label{e11}; \\
& \begin{pmatrix} 1 \\ 1 \\ 1 \\ \end{pmatrix}, & ~~\begin{pmatrix}  1  \\  \omega  \\  \omega^{\ast}  \\ \end{pmatrix},~~\begin{pmatrix}  1 \\  \omega^{\ast} \\ \omega \\ \end{pmatrix} \label{e12}; \\
& \begin{pmatrix} 1 \\ \omega \\ 1 \\ \end{pmatrix}, & ~~\begin{pmatrix}  1 \\  \omega^{\ast} \\ \omega^{\ast} \\ \end{pmatrix}, ~~\begin{pmatrix}  1  \\  1  \\  \omega  \\ \end{pmatrix} \label{e13}; \\
& \begin{pmatrix} 1 \\ \omega^{\ast} \\ 1 \\ \end{pmatrix}, & ~~\begin{pmatrix}  1  \\  1  \\  \omega^{\ast}  \\ \end{pmatrix},~~\begin{pmatrix}  1 \\  \omega \\ \omega \\ \end{pmatrix}, \label{e14}
\end{eqnarray}
where $\omega=e^{i\frac{2\pi}{3}}$ and $~^{\ast}$ stands for the complex conjugation. Note, that the second basis (\ref{e12}) is the Fourier transform of the computational basis. If the basis (\ref{e11}) is the eigenbasis of $S_z$, then the operator $exp(-iS_z t)=\text{diag}\{1,e^{-it},e^{-i2t}\}$
generates a permutation inside the bases (\ref{e12}-\ref{e14}) for times being the multiple of $2\pi/3$. This is an analogy to the $\pi$ rotation about $z$ axis for spin $1/2$ which generates the swap operation in the $S_x$ and $S_y$ bases. The transition between different bases, other than the computational one, can be obtained by the operator $\text{diag}\{1,\omega,1\}$ generated by the Hamiltonian proportional to $S_z^2$ which is the one-axis twisting squeezing generator \cite{KU}. 

\begin{figure}
\scalebox{1.0}  {\includegraphics[width=8truecm]{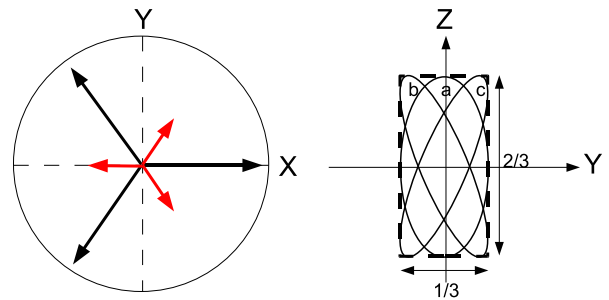}}
\caption{\label{f1} Left: mean spin vectors of the bases (\ref{e12}) black and (\ref{e13},\ref{e14}) red. Right: the squeezing of states (\ref{e12}-\ref{e14})} --- a,b and c respectively.
\end{figure}

The states in the first column of the bases (\ref{e12}-\ref{e14}) are of the form (\ref{e5}), therefore the corresponding mean spin vectors point in $x$ direction and due to the rotational symmetry the other states from these bases have the same properties. The lengths of the mean spin vectors of the MUBs (\ref{e12}-\ref{e14}) are $2\sqrt{2}/3$, $\sqrt{2}/3$ and $\sqrt{2}/3$ respectively. The vectors lie in the $xy$ plane and point in the directions $0$, $2\pi/3$ and $-2\pi/3$ for the basis (\ref{e12}) and in the directions $\pi$,$\pi/3$ and $-\pi/3$ for the bases (\ref{e13}) and (\ref{e14}). One may find that in general the shorter the mean spin vector is, the more squeezed the corresponding state has to be. On the other hand, Eq. (\ref{e8}) states that for all states of the form (\ref{e5}) the uncertainties $\Delta S_z^2$ and $\Delta S_y^2$ do not depend on the state, suggesting that all these states are equally squeezed what seems in contrary to the fact that vectors (\ref{e13}) and (\ref{e14}) are shorter than vectors (\ref{e12}).  However, it may happen that the states with the shorter mean spin may be squeezed in the direction other than $y$, but still have the same uncertainties $\Delta S_z^2$ and $\Delta S_y^2$. This idea is depicted in the Fig. {\ref{f1}} right.
Indeed, by studying the rotation about $x$ axis one may show that the states lying along $x$ axis are squeezed in the $y$ direction, the direction tilted by an angle $\approx-\frac{\pi}{6}$ and $\approx\frac{\pi}{6}$ for the bases (\ref{e12}-\ref{e14}) respectively. The uncertainties in the direction of squeezing for the last two bases reaches slightly bellow $0.06$. Note that the bases (\ref{e13}) and (\ref{e14}) seems similar and the complex conjugation changes one basis to the other. The conjugation applied to the spin vector 
$\vec{S}=\{S_x,S_y,S_z\}$ generates reflection in the $xz$ plane inverting $y$ axis $\vec{S^{\ast}}=\{S_x,-S_y,S_z\}$. This is visible in the uncertainties and in the distribution of the mean spin vectors (see Fig. \ref{f1}).

\begin{figure}
\scalebox{1.0}  {\includegraphics[width=4truecm]{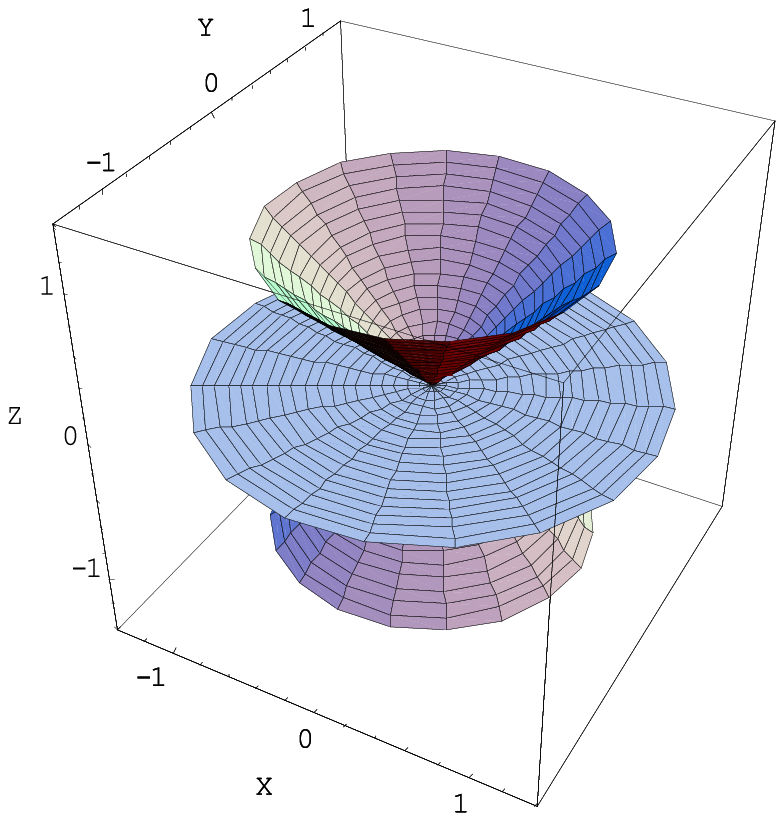}}{\includegraphics[width=4truecm]{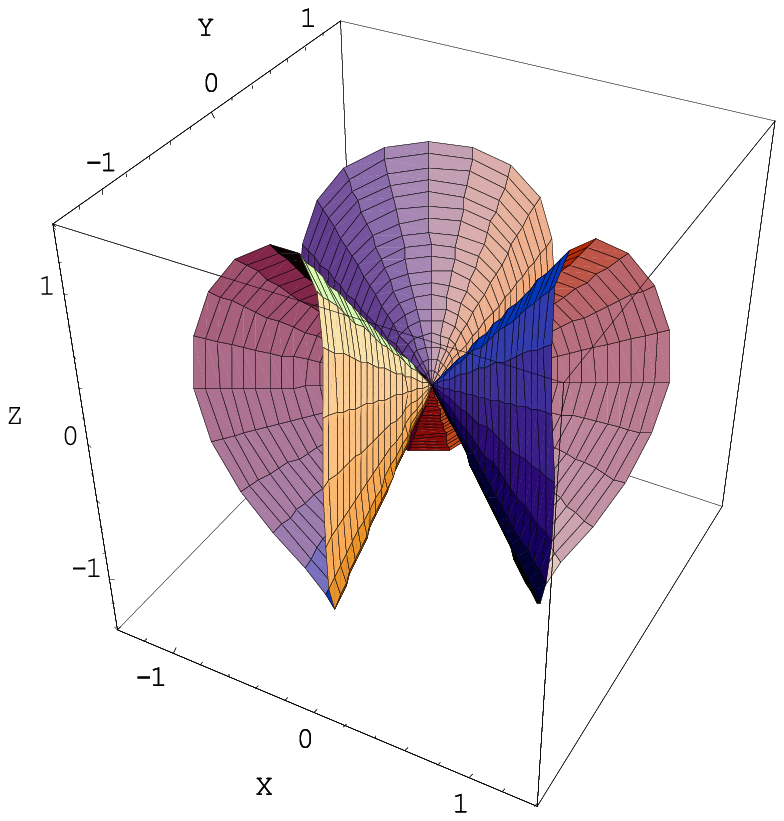}}
\caption{\label{f2} Graphical representation of states corresponding to MUBs (\ref{e11}) left and (\ref{e12}) right.}
\end{figure}

Spin states are sometimes depicted as the cones in the three dimensional space, where the surface of the cone represents an area covered by the spin vector due to its spread caused by the uncertainty principle and its dilation angle represents the uncertainties. In the Fig. \ref{f2} we suggest how one may picture spin squeezed states corresponding to the MUBs (\ref{e11}) and (\ref{e12}). 

\section{Complementary Observables}

In this section we are going to look for the physical meaning of the observables corresponding to spin 1 MUBs. The MUBs were taken to be the eigenstates of the four unitary operators from the Weyl-Heisenberg group $U$, $V$, $UV$ and $UV^2$ where the operator $U$ is the permutation operator of the computational basis and $V$ is the diagonal operator causing the phase shift. All four operators are non-degenerate and their eigenvalues are the third roots of unity. One may interpret the generators of $U$ and $V$ as momentum and position operators respectively acting on the discrete space with only three allowed positions and periodic boundary conditions. The eigenbasis of $V$, the computational basis, has been chosen to be the eigenbasis of $S_z$ operator, therefore this MUB corresponds to the knowledge of the spin projection onto $z$ axis and one may think of it as some kind of position operator. The second MUB, the Fourier transform of the computational basis, might be interpreted as an analog of momentum operator. This kind of operator causes the cyclic permutation of eigenstates of $S_z$ $$|S_z=-1\rangle\rightarrow|S_z=0\rangle\rightarrow|S_z=1\rangle\rightarrow|S_z=-1\rangle.$$
However, from the physical point of view, this permutation is abstract and hard to interpret since it is not a simple spin rotation, because it takes one coherent state to the null projection state, then to the other coherent state and next back to the initial coherent state --- it has to be a nontrivial combination of squeezing and rotation. The same problem remains for the other two MUBs. The states of the MUBs (\ref{e12}-\ref{e14}) are bizarre in the sense that they are neither coherent, nor maximally squeezed null projection states and the physical meaning of the corresponding observables is not as simple as of the standard spin projection operator $S_z$. One may wander whether it would be more convenient to choose a different computational basis.

\begin{figure}
\scalebox{1.0}  {\includegraphics[width=4truecm]{fig1.eps}}{\includegraphics[width=4truecm]{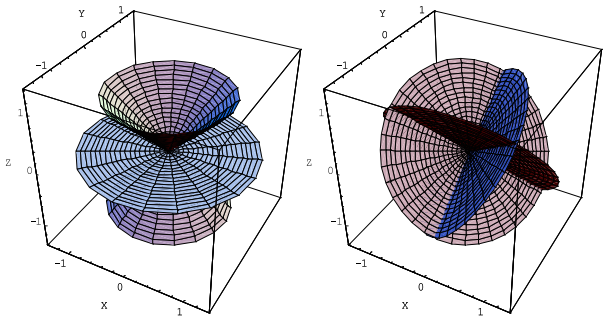}}
\caption{\label{f3} Graphical representation of two MUBs: left, the basis (\ref{e11}) and right, an example of basis made of states (\ref{e6}).}
\end{figure}

Among all the states unbiased to the $S_z$ basis there is a class of maximally squeezed states given by Eq. (\ref{e6}) corresponding to the null projection states onto all axes tetrahedral to $z$. One may find that in this class there are infinitely many ways to pick a three mutually orthogonal states forming up a basis which is mutually unbiased to the $S_z$ one. These states might be represented in the real space as three mutually orthogonal planes (see Fig. {\ref{f3}}).
In general, any three states corresponding to the null projection states onto a three arbitrary mutually orthogonal directions form up a basis in the Hilbert space of spin 1. Let us choose as a computational basis a basis made of a three null projection states along $x$, $y$ and $z$ directions, which in the $S_z$ basis are given by:
$$|x\rangle=\begin{pmatrix} \frac{1}{\sqrt{2}} \\ 0 \\ - \frac{1}{\sqrt{2}} \\ \end{pmatrix},~~|y\rangle=\begin{pmatrix}  \frac{i}{\sqrt{2}} \\ 0 \\  \frac{i}{\sqrt{2}} \\ \end{pmatrix},~~|z\rangle=\begin{pmatrix} 0 \\ 1 \\ 0 \\ \end{pmatrix}.$$
Usually, the state $|y\rangle$ is written as $(\frac{1}{\sqrt{2}},0, \frac{1}{\sqrt{2}})^T$, but there is a reason why we multiplied it by $i$. Any state of the form
\begin{equation}\label{e15}
|\theta,\varphi\rangle=\sin\theta\cos\varphi|x\rangle+
\sin\theta\sin\varphi|y\rangle+\cos\theta|z\rangle
\end{equation}
is also a null projection state along direction $\vec{n}=(\sin\theta\cos\varphi,\sin\theta\sin\varphi,\cos\theta)$, therefore all real linear combinations of $|x\rangle$, $|y\rangle$ and $|z\rangle$ resemble the vectors in the Euclidean space $\mathbb{R}^3$. What is interesting, the Fourier transform of the new computational basis is the eigenbasis of the spin projection operator $S_m$ along direction $\vec{m}=(1,-1,-1)$ which is tetrahedral to z. Moreover, the operator $S_m$ generates a rotation about $\vec{m}$, which for an angle being a multiple of $2\pi/3$ causes a cyclic permutation of basis states --- the $2\pi/3$ rotation about $\vec{m}$ transforms $xy$ plane into $xz$ plane, etc. This means that the projection operator $S_m$ might be considered as the {\it momentum} operator $P$ in the chosen basis. Actually, it could be any projection operator along one of four tetrahedral directions $(1,\pm 1,\pm 1)$.

What is the corresponding {\it position} operator? It has to be an operator with the eigenvectors $|x\rangle$, $|y\rangle$ and $|z\rangle$. In general, we are looking for the operator whose eigenvectors are the null projection states along a three mutually orthogonal directions $i$, $j$ and $k$. It happens that this is the two-axis countertwisting squeezing operator \cite{KU} of the form $S_i^2-S_j^2$ with the eigenvalues 1, 0 and $-1$. It might be as well represented as $S_i^2-S_k^2$ or $S_j^2-S_k^2$, because different representation does not change the eigenvectors, but the distribution of the eigenvalues, thus the {\it position} operator could be taken as $X=S_x^2-S_y^2$. Next, let us find what information do we gain while measuring $X$. The information related to measuring $|j\rangle$ is that the spin is definitely not lying along direction $j$, therefore the measurement of $X$ gives an answer to the question: {\it Along which one of the three mutually orthogonal axes the spin is not lying?} Note, that the uncertainty principle forbids the spin to lie definitely along one axis, that is why the question we can ask sounds a little bit odd --- in quantum world we are not allowed to ask about the direction in which the spin is pointing. Eventually, the two complementary spin 1 observables might be reformulated as the two complementary questions:
\begin{itemize}
\item {\it Along which one of the three mutually orthogonal axes $i$, $j$ and $k$ the spin is not lying?}
\item {\it What is the spin projection onto one of the four axes tetrahedral to $i$, $j$ and $k$?}
\end{itemize} 
Having an answer to the one of the above questions one knows nothing about an answer to the other question.

Even that we have chosen different computational basis, the remaining two MUBs (\ref{e13}) and (\ref{e14}) are squeezed, but not maximally squeezed, i.e. they are not the null projection states. The corresponding observables, which might be taken as a real linear combinations of a projectors $|\psi\rangle\langle\psi|$ onto MUBs states, are a generators of a transformations which are a combination of a rotation and a squeezing, what makes them hard to identify as some simple physical quantities or to associate them with reasonable questions one can ask about spin. However, it would be very interesting to find the physical meaning of the observables whose eigenstates are partially squeezed and we leave this problem as an open question.

\section{Generation and Measurement of Spin 1 MUBs}

MUBs have found many practical applications in quantum information processing. In order to prepare an unbiased states and to perform a suitable measurements one has to know how to transform between the different MUBs. In the case of spin 1 some states are easier to obtain as well as some measurements are easier to perform. In order to implement a certain information processing tasks, the ability of preparing and measuring all possible states is desired --- we need spin 1 to be an exact implementation of a qutrit. For the study of the possibility of spin 1 implementation of a qutrit see \cite{Das}. The most common and easiest measurements of spin are the one of the Stern-Gerlach type, although generalized Stern-Gerlach measurements have been also proposed \cite{SW}. The prior measurement could be also considered as a preparation procedure, therefore the coherent states and the maximally squeezed null projection states are the most accessible ones. It is also obvious, that the most natural choice of the computational basis should be the eigenbasis of $S_z$ for some reference axis $z$. The transformation between the different coherent states is relatively easy, since it requires only a spin rotation via an application of a linear magnetic field. However, the transformation between a coherent and a null projection states is no longer simple, because it cannot be obtained by a linear effects. 

We already mentioned that if the computational basis corresponds to $S_z$, the transformation within the three remaining MUBs could be obtained by a rotation about $z$. It is simply an analog of translation or boost. On the other hand, the transformation between these MUBs is done via one axis twisting squeezing. This kind of squeezing is generated by operator $exp(-iS_z^2t)$ and the change between the MUBs occurs for the times being the multiple of $2\pi/3$. The above operator is of course nonlinear and requires a quadrupole effects. Still, the most important question is how to obtain the three basis from the $S_z$ basis. The usual way of generating the MUBs states is to perform the Fourier transform on the computational basis. Let us look for a more general transformation, a Fourier like transform, taking the basis state to the states (\ref{e2}). This transformation is represented by the unitary matrix with all entries having an equal absolute value, a complex Hadamard matrix. For the spin 1 such a transformation might be obtained for the one-axis twisting about an axis tetrahedral to $z$ whose direction is given by Eq. (\ref{e10}). The corresponding operator $exp(-iS_{m_{\varphi}}^2t)$ becomes a Hadamard for $t=2\pi/3$ (and $4\pi/3$) which up to a global phase is given by
\begin{equation}\label{e16}
\frac{1}{\sqrt{3}} \begin{pmatrix} 
e^{i\frac{\pi}{3}} & e^{-i\varphi} & e^{-i2\varphi} \\ 
e^{i\varphi} & e^{i\frac{\pi}{3}} & -e^{-i\varphi}\\ 
e^{i2\varphi} & -e^{i\varphi} & e^{i\frac{\pi}{3}} \\ 
\end{pmatrix}.
\end{equation}
Straightforward calculations show that the above operation generates three squeezed states symmetrically distributed by an angle $\frac{2\pi}{3}$ on the $xy$ plane whose lengths do not depend on $\varphi$ and are equal $\sqrt{2/3}$. The angle $\varphi$ is only affecting the global deviation of the basis from the alignment along $0$, $\frac{2\pi}{3}$ and $-\frac{2\pi}{3}$. In order to obtain the full Fourier transform one still has to apply the one-axis twisting and a rotation, both about $z$. Even without this, the operator (\ref{e16}) generates a basis which is unbiased to $S_z$, what is enough to perform certain tasks like quantum cryptography on a three level system. The other two bases might be obtained by the one-axis twisting $\pm 2\pi/3$ pulse about $z$.

To perform the measurement in a basis corresponding to a certain MUB one needs only to know the procedure of transforming this MUB into the computational basis. In this case one or two one-axis twisting $2\pi/3$ pulses about $z$ has to be followed by the inverse of (\ref{e16}). Then, it is enough to perform a Stern-Gerlach measurement in the computational basis. Eventually, the $\pm 2\pi/3$ rotation and $\pm 2\pi/3$ one-axis twisting pulses about $z$, together with the Fourier like transform (\ref{e16}), its inverse and the Stern-Gerlach measurement gives the full set of operations needed to perform three level quantum cryptography or the tomography of spin 1.

\section{Higher Spin Numbers}

One may also consider the MUBs for the higher spin numbers, although for some $s$ we do not even know how many MUBs there exist \cite{Prob}. Because of this fact, in this section we only consider the two MUBs, the eigenbasis of $S_z$ and its Fourier transform. The Hilbert space is $d=2s+1$ dimensional. Again, the mean spin vectors of all the states unbiased to $S_z$ basis are lying in the $xy$ plane, due to the similar reason as before (recall Eq. (\ref{e3})), and the $k 2\pi/d$ rotation about $z$, k being an integer, causes the translation between the states of the Fourier transform basis. Once more, because of the rotational symmetry we may narrow our study to the state
\begin{equation}\label{e17}
\frac{1}{\sqrt{d}} \begin{pmatrix} 
1 \\ 1 \\ \vdots \\ 1 \\
\end{pmatrix}.
\end{equation}
The mean spin vector corresponding to the above state has to point in the $x$ direction, because for all $d$ $\langle S_y \rangle=0$. This is because $S_y$ is an antisymmetric matrix and an expectation value of any matrix with respect to the state (\ref{e17}) equals the sum of all matrix elements divided by $d$. The length of the mean spin vector is given by
\begin{equation}\label{e18}
\langle S_x \rangle = \frac{1}{d}\left(\sum_{j=1}^{d-1}\sqrt{j(d-j)}\right)< \frac{d-1}{2}=s.
\end{equation}
It is shorter than the length of the mean spin of a coherent state what indicates that the state (\ref{e17}) is squeezed, as well as the other Fourier transform states have to be. Indeed,
\begin{equation}\label{e19}
\Delta S_z^2=\frac{d^2-1}{12}=\frac{s(s+1)}{3} > \frac{s}{2}, 
\end{equation}
but 
\begin{eqnarray}
\Delta S_y^2 &=& \sum_{j=1}^{d-1}\frac{j(d-j)}{2d}-\sum_{j=1}^{d-2}\frac{\sqrt{j(d-j)(j+1)(d-j-1)}}{2d}  \nonumber \\
&<&\sum_{j=1}^{d-1}\frac{j(d-j)}{2d}-\sum_{j=1}^{d-2}\frac{j(d-j-1)}{2d}= \nonumber \\
&=&\frac{d-1}{2d}+\sum_{j=1}^{d-2}\frac{j}{2d}=\frac{d-1}{4}=\frac{s}{2}, \label{e20}
\end{eqnarray}
therefore the state has to be squeezed.

Once more, it is hard to interpret the physical meaning of the operator corresponding to the Fourier transform of the $S_z$ basis because, as in the case of spin 1, these states are neither coherent nor null-projection states. It is also hard to tell whether one may find an analogs of momentum and position operators, with one of them being simply the spin projection operator onto some direction. 

\section{Conclusions}

We have discussed the physical aspects of spin 1 complementary observables, introducing the analogs of position and momentum operators and showing that the spin squeezed states can play an important role in the MUBs studies. Moreover, we proposed the methods of generation and measurement of the different spin 1 MUBs what is necessary for spin 1 quantum information processing. In the recent years the biggest experimental progress in quantum information had happened within quantum optics and it is important to mention how our work is related to this field. One of the most promising optical qutrit implementations is the polarization states of a biphoton --- the joint polarization states of a two indistinguishable photons. Still, even for biphoton, the generation of all possible states requires nonlinearities from either nonlinear crystals \cite{Bog1,Bog2} or measurement-induced state filtering \cite{Lan}. This requirement resemble the quadrupole nonlinearity needed to obtain the spin squeezed states. It seems that the nature truly reveals its quantum behavior through the nonlinear effects. 

The state of spin 1 may be represented as a product state of two 1/2 spins. The Hilbert space of spin 1/2 is isomorphic to the photon polarization space, thus the product representation of spin 1 is isomorphic to the two photon product polarization representation. Recently, a two qutrit cryptographic protocols have been designed for the biphoton \cite{Bre}, and one of them, the so called umbrella protocol, relies on a two MUBs. It is easy to see that the two MUBs of the umbrella protocol correspond to the MUBs of the analogs of position and momentum operators. One of them consists only of maximally entangled photon states, this is our basis made of null projection states, while the other one, tetrahedral to the first one, consists of the two non-entangled product states and the one maximally entangled state, which in our case is spin projection basis made of the two coherent states and the one null projection state. The authors of the protocol have shown that even using only the two of the four MUBs one can achieve greater efficiency than using the three MUB qubit protocol. Our results show that the umbrella protocol can be implemented on spin 1.
 
We would like to thank Genowefa {\'S}l{\'o}sarek for the discussions on spin 1 tomography. P.K. acknowledges the support from Tomasz {\L}uczak from subsidium MISTRZ (Foundation for Polish Science) and the Scientific Scholarship of the City of Pozna{\'n}.

\end{document}